\documentclass[twocolumn]{aastex62}

\usepackage{natbib}
\usepackage{amsmath}
\usepackage{subfigure}
\setcitestyle{sort,comma}
\usepackage{csvsimple}

\graphicspath{{./}{figures/}}

\shorttitle{Water in the transmission spectrum of GJ~9827\,d}
\shortauthors{}

\begin{document}

% Special commands

\title{\large{Water absorption in the transmission spectrum of the water-world candidate GJ~9827\,d}}

\correspondingauthor{Pierre-Alexis Roy}
\email{pierre-alexis.roy@umontreal.ca}

\author[0000-0001-6809-3520]{Pierre-Alexis Roy} 
\affil{Department of Physics and Trottier Institute for Research on Exoplanets, Universit\'{e} de Montr\'{e}al, Montreal, QC, Canada}

\author[0000-0001-5578-1498]{Bj\"{o}rn Benneke} 
\affil{Department of Physics and Trottier Institute for Research on Exoplanets, Universit\'{e} de Montr\'{e}al, Montreal, QC, Canada}

\author[0000-0002-2875-917X]{Caroline Piaulet}
\affil{Department of Physics and Trottier Institute for Research on Exoplanets, Universit\'{e} de Montr\'{e}al, Montreal, QC, Canada}

\author[0000-0002-4020-3457]{Michael A. Gully-Santiago}
\affil{Department of Astronomy, University of Texas at Austin, Austin, TX 78712, USA}

\author{Ian J.M. Crossfield} 
\affil{The University of Kansas, Department of Physics and Astronomy, Malott Room 1082, 1251 Wescoe Hall Drive, Lawrence, KS, 66045, USA}

\author[0000-0002-4404-0456]{Caroline V. Morley}
\affil{Department of Astronomy, University of Texas at Austin, Austin, TX 78712, USA}

\author[0000-0003-0514-1147]{Laura Kreidberg} 
\affil{Max Planck Institute for Astronomy, K\"{o}nigstuhl 17, D-69117 Heidelberg, Germany}
\affil{Center for Astrophysics, Harvard \& Smithsonian, 60 Garden Street, Cambridge, MA, 02138, USA}

\author[0000-0001-5442-1300]{Thomas Mikal-Evans}
\affil{Max Planck Institute for Astronomy, K\"{o}nigstuhl 17, D-69117 Heidelberg, Germany}

\author[0000-0002-2072-6541]{Jonathan Brande}
\affil{Department of Physics and Astronomy, University of Kansas, 1082 Malott, 1251 Wescoe Hall Dr., Lawrence, KS 66045, USA}

\author[0000-0002-4771-0312]{Simon Delisle}
\affil{Department of Physics and Trottier Institute for Research on Exoplanets, Universit\'{e} de Montr\'{e}al, Montreal, QC, Canada}

\author[0000-0002-8963-8056]{Thomas P. Greene}
\affil{Space Science and Astrobiology Division, NASA Ames Research Center, MS 245-6, Moffett Field, CA 94035, USA}

\author[0000-0003-3702-0382]{Kevin K.\ Hardegree-Ullman}
\affiliation{Steward Observatory, The University of Arizona, Tucson, AZ 85721, USA}

\author[0000-0002-7129-3002]{Travis Barman}
\affil{Lunar and Planetary Laboratory, The University of Arizona, 1640 E. University Blvd, Tucson, AZ 85721, USA}

\author[0000-0002-8035-4778]{Jessie L. Christiansen}
\affil{Caltech/IPAC, M/S 100-22, 1200 E. California Blvd., Pasadena, CA 91125, USA}

\author[0000-0003-2313-467X]{Diana Dragomir}
\affil{Department of Physics and Astronomy, University of New Mexico, 1919 Lomas Blvd. NE, Albuquerque, NM 87131, USA}

\author[0000-0002-9843-4354]{Jonathan J. Fortney}
\affil{Department of Astronomy \& Astrophysics, University of California, Santa Cruz, CA 95064, USA}

\author[0000-0001-8638-0320]{Andrew W. Howard}
\affil{Department of Astronomy, California Institute of Technology, Pasadena, CA 91125, USA}

\author[0000-0002-6115-4359]{Molly R. Kosiarek}
\affil{Department of Astronomy and Astrophysics, University of California, Santa Cruz, CA 95064, USA}

\author[0000-0003-3667-8633]{Joshua D. Lothringer}
\affil{Department of Physics and Astronomy, Johns Hopkins University, Baltimore, MD 21210, USA}

\begin{abstract}
% Abstract - March 2023
% Basic introduction to the field (1-2)
Recent work on the characterization of small exoplanets has allowed us to accumulate growing evidence that the sub-Neptunes with radii greater than $\sim2.5\,R_\oplus$ often host H$_2$/He-dominated atmospheres both from measurements of their low bulk densities and direct detections of their low mean-molecular-mass atmospheres.
% More detailed background (2-3)/ The problem (1)
However, the smaller sub-Neptunes in the 1.5-2.2 R$_\oplus$ size regime are much less understood, and often have bulk densities that can be explained either by the H$_2$/He-rich scenario, or by a volatile-dominated composition known as the ``water world" scenario.
% Here we report (1)
Here, we report the detection of water vapor in the transmission spectrum of the $1.96\pm0.08$ R$_\oplus$ sub-Neptune GJ~9827\,d obtained with the Hubble Space Telescope.
% Main results (2-3)
We observed 11 HST/WFC3 transits of GJ~9827\,d and find an absorption feature at 1.4$\mu$m in its transit spectrum, which is best explained (at 3.39$\sigma$) by the presence of water in GJ~9827\,d's atmosphere. We further show that this feature cannot be caused by unnoculted star spots during the transits by combining an analysis of the K2 photometry and transit light-source effect retrievals. We reveal that the water absorption feature can be similarly well explained by a small amount of water vapor in a cloudy H$_2$/He atmosphere, or by a water vapor envelope on GJ~9827\,d.   
% Some more perspective (2-3)
Given that recent studies have inferred an important mass-loss rate ($>0.5\,$M$_\oplus$/Gyr) for GJ~9827\,d making it unlikely to retain a H-dominated envelope, our findings highlight GJ~9827\,d as a promising water world candidate that could host a volatile-dominated atmosphere. This water detection also makes GJ~9827\,d the smallest exoplanet with an atmospheric molecular detection to date.

\end{abstract}


% \tableofcontents

%Paragraph

\section{Introduction} \label{sec:intro}
% Paragraph 1 : Sub-Neptunes and water worlds
While many questions remain regarding the nature of sub-Neptune exoplanets, the last decade of transmission spectroscopy with the Hubble Space Telescope (HST) has shown that the larger sub-Neptunes are often best described by H$_2$/He-dominated atmospheres \citep[e.g., ][]{benneke_sub-neptune_2019, benneke_water_2019, mikal-evans_transmission_2020, kreidberg_tentative_2022}. However, this picture is much less clear when considering sub-Neptunes that are in the smaller 1.5-2.2 R$_\oplus$ size-regime, near the radius valley \citep{fulton_california-_2017, fulton_california-_2018, vaneylen_asteroseismic_2018, hardegree-ullman_scaling_2020}. These planets have bulk densities than can be explained by either the H$_2$/He-rich sub-Neptune scenario, or by a volatile-dominated composition, where water (or another molecule of similar mean-molecular-weight) supplants H$_2$ and He as the most abundant atmospheric species \citep{rogers_three_2010, luque_density_2022, rogers_conclusive_2023}. This type of exoplanet has been long theorized and is referred to as ``water world" \citep{adams_ocean_2008, acuna_water_2022}. 

% Paragraph 2: Water worlds 
These smaller sub-Neptunes, which are often inconsistent with extended H-dominated atmospheres with large scale heights \citep{aguichine_massradius_2021, piaulet_evidence_2022}, are also found in a smaller mass regime than the larger sub-Neptunes, making these close-in planets much more exposed to mass-loss processes \citep{owen_atmospheric_2019}, and thus more likely to have lost their H$_2$ and He envelope over their lifetime. A recent study found a first line of evidence for the existence of such volatile-rich water worlds in the super-Earth Kepler-138\,d, by combining a thorough interior analysis of the planet with mass-loss estimates, effectively showing that this super-Earth cannot be purely rocky, but that it also cannot retain a hydrogen layer \citep{piaulet_evidence_2022}. However, the direct spectroscopic confirmation of a volatile-rich high mean-molecular-weight atmosphere on a water world candidate still eludes us, and such a result would provide a new line of evidence for the water worlds.

%Paragraph 3: Introduce the planet -  Sub-Neptune Characterizable in transit
The discovery of the transiting sub-Neptune GJ~9827\,d \citep{niraula_three_2017, rodriguez_system_2018} represents a rich opportunity to characterize the atmosphere of a warm sub-Neptune via transmission spectroscopy and to deepen our understanding of this potential water world \citep{aguichine_massradius_2021}. Rapidly orbiting (6.2 days) a low-mass K6V star with a size of $1.96\pm0.08\,R_\oplus$, a mass of $3.4\pm0.6\,M_\oplus$ \citep{kosiarek_physical_2021}, and a zero-albedo equilibrium temperature of $680 \pm 25\,$K \citep{rodriguez_system_2018}, GJ~9827\,d allows us to obtain a high signal-to-noise ratio (S/N) in transmission spectroscopy, and to add a precious new target to the sample of sub-Neptunes with transit spectra. While JWST now allows to observe the eclipses and phase curves of small exoplanets deeper in the infrared \citep[e.g.,][]{kempton_reflective_2023}, transit spectroscopy remains the best method to obtain in-depth looks into the atmospheres of sub-Neptunes and potential water worlds with HST, as they rarely are hot enough to provide high S/N in the near-infrared \citep[hot Neptune desert;][]{owen_photoevaporation_2018}.

% Removed for length purposes
%Paragraph 3: GJ~9827d is special as a intriguing almost-keystone planet in terms of formationéradius valley studies
% As the outer-most planet of a close-in, near-resonant, three-planet system, GJ~9827\,d also represents a ripe opportunity to study planet formation theories. GJ~9827\,d finds itself right on the edge of the low-mass-star radius valley \citep{cloutier_evolution_2020}, and thus represents a target of particular interest for studying the transition between gaseous and rocky planets, particularly since the two inner super-Earths are on the opposite ``rocky" side of the radius valley.

%Paragraph 4: Predicted interior composition, etc. (shorter)
While the average density of GJ~9827\,d has now been constrained ($>3\sigma$) by numerous RV studies of the system \citep{prieto-arranz_mass_2018, rice_masses_2019, kosiarek_physical_2021}, there still remains ambiguity regarding its bulk composition, as its density could be explained by a range of compositions from an extended H$_2$/He layer to a water world composition with a $\sim$20\,\% water mass fraction \citep{aguichine_massradius_2021}. However, the high irradiation of the planet, the old age of the system \citep{kosiarek_physical_2021} and the non-detections of HeI and H$\alpha$ absorption from the ground \citep{kasper_nondetection_2020, carleo_multiwavelength_2021, krishnamurthy_absence_2023} make it unlikely that GJ~9827\,d would have retained a primordial H-dominated envelope to date.

In this work, we present the most precise look yet at GJ~9827\,d via transmission spectroscopy with the Hubble Space Telescope Wide Field Camera 3 (HST/WFC3), and reveal a water absorption feature in its transmission spectrum. In Section 2, we present the observations obtained for this study and we describe the data analysis in Section 3. Section 4 presents our atmospheric analysis of the HST transit spectrum and the related results are presented in Section 5. We end by discussing our findings and presenting our conclusions in Section 6.

\begin{figure*}[t!]
\begin{center}
\includegraphics[width=0.95\linewidth]{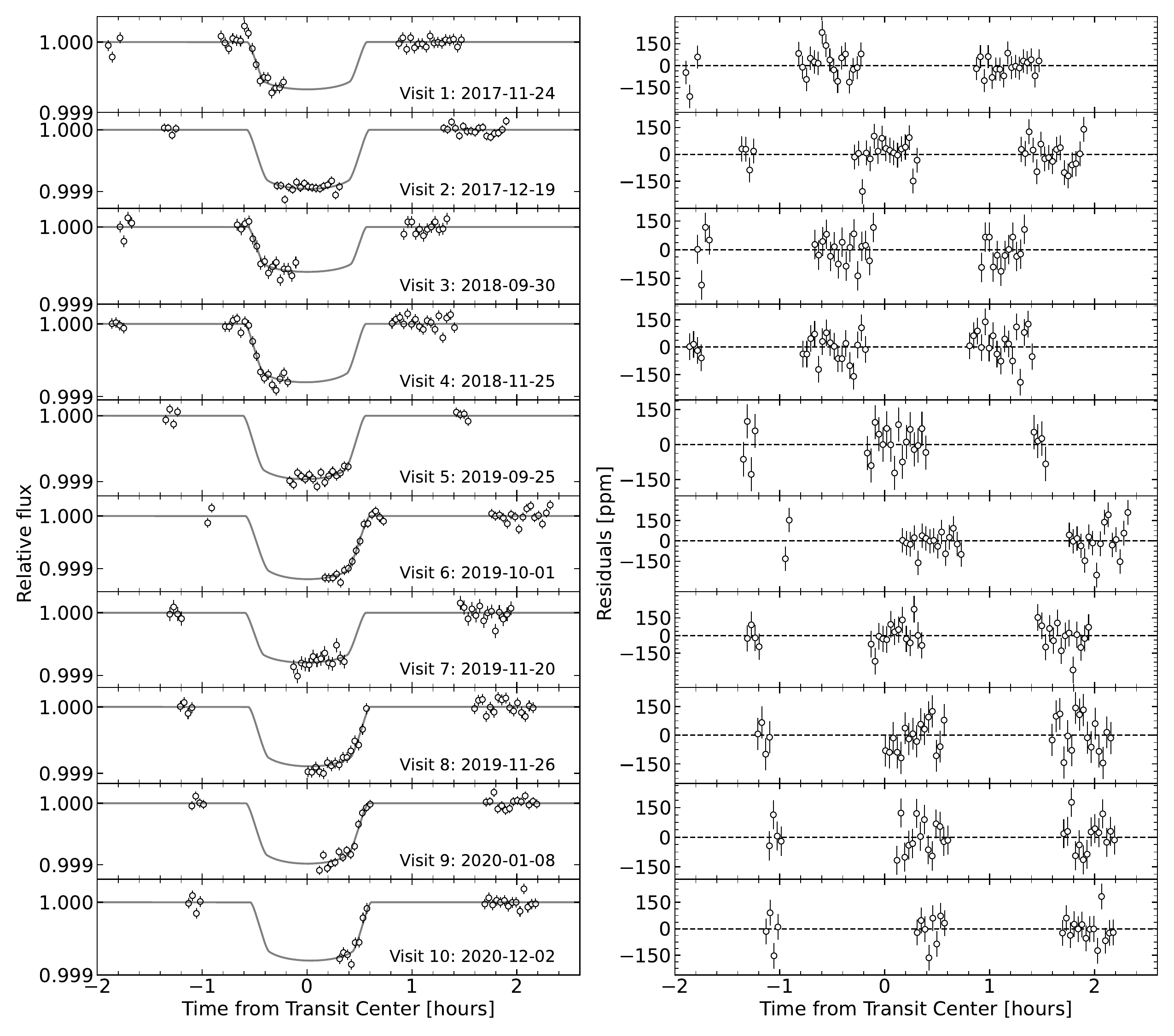}
\end{center}
\caption{All 10 HST/WFC3 broadband light-curve fits of the transits of GJ~9827\,d. \textbf{Left:} Systematics-corrected and normalized broadband light curves for the 10 transits of GJ~9827\,d (data points). Each visit is centered around the fitted transit time for that visit. The best-fitting transit model is also shown as the grey line. \textbf{Right:} Residuals of the broadband light-curve fits shown on the left.}
\label{fig:wlc}
\end{figure*}

\section{Observations and data reduction}\label{sec:obs_red}
GJ~9827\,d was observed transiting its host star 11 times between December 2017 and December 2020 with the Wide Field Camera 3 (WFC3) onboard the Hubble Space Telescope (HST) as part of the mini-Neptune atmosphere diversity survey (GO 15333: PI Crossfield). The G141 grism was used in order to obtain the transmission spectrum of the sub-Neptune over the 1.1-1.7 $\mu$m range. %The large number of transits was chosen to increase the SNR and precision of our transit spectrum and enable us to robustly detect (or rule-out) potential molecular absorption features in the atmosphere of the warm sub-Neptune. 
%The short 6.2\,d period of GJ~9827\,d's orbit enabled us to observe all 11 transits within $\sim$3~yr.

\begin{figure}[t!]
  \centering
    \includegraphics[width=\linewidth]{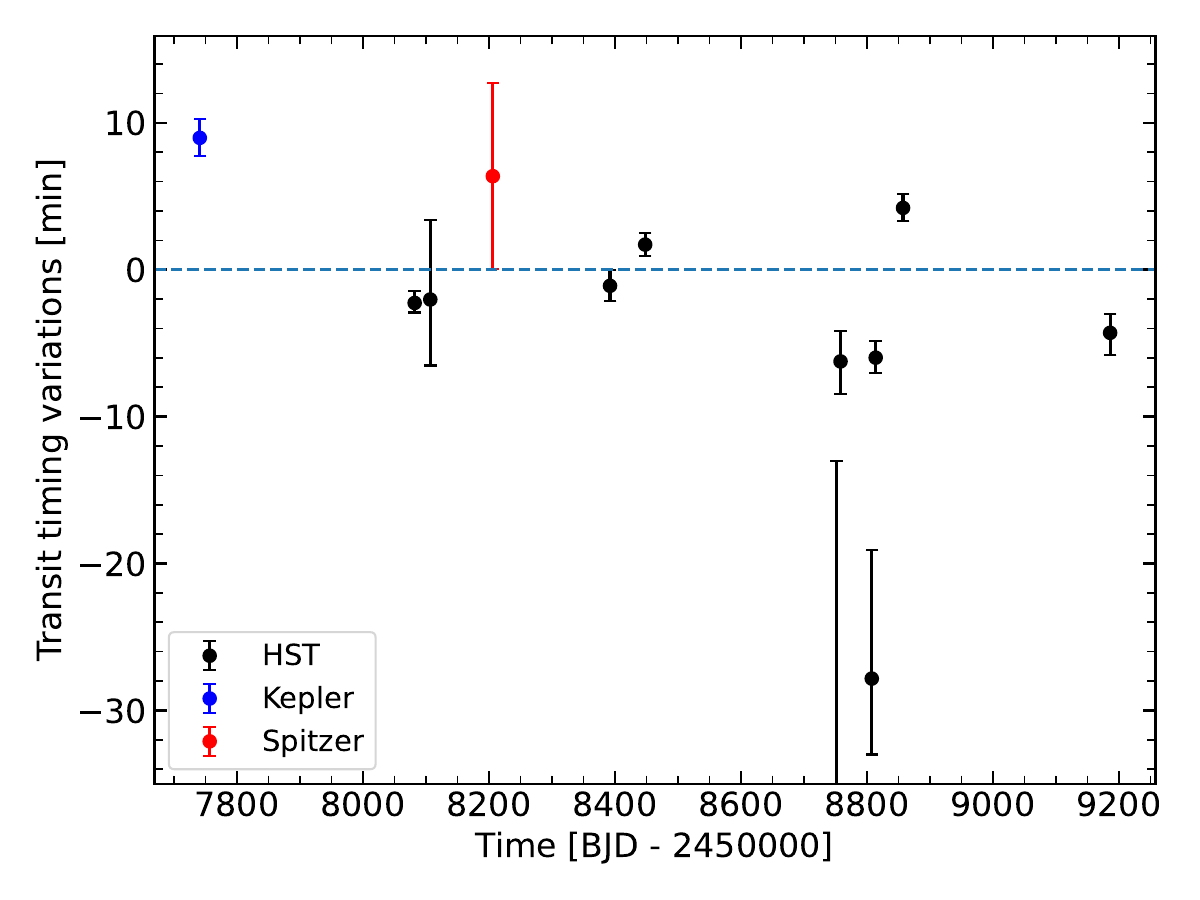}
  \caption{Observed mid-transit time (black points) for each of the 10 transits compared to a fitted linear ephemeris to all transit timings presented (except the two timings with large uncertainties; $T_{0, BJD} = 2459185.987 \pm 0.002$ , $P = 6.20186 \pm 0.00002$ d) . The ephemeris derived from the K2 campaign is shown \citep[blue;][]{niraula_three_2017} along with the Spitzer transit for this planet \citep[red;][]{kosiarek_physical_2021}. Our 10 HST/WFC3 transits suggest that there are statistically significant TTVs in the orbit of GJ~9827\,d of the order of 5-10 minutes.}
  \label{fig:TTVs}
\end{figure}

\begin{figure*}[t!]
\begin{center}
\includegraphics[width=0.92\linewidth]{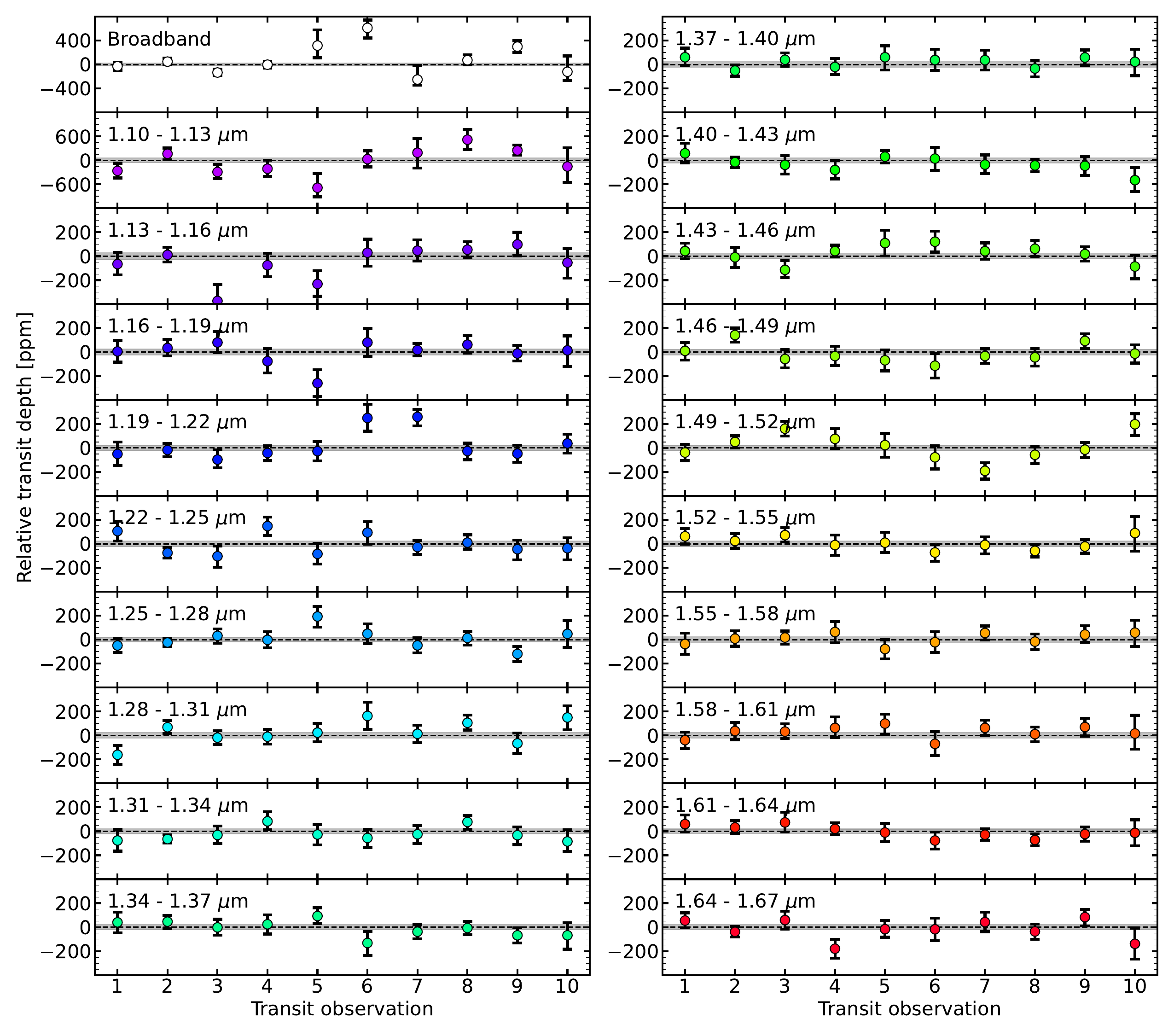}
\end{center}
\caption{Individual relative transit depths compared to the relative final transmission spectrum. For each of the 10 transmission spectra, we obtain the relative transit spectrum by subtracting the average across wavelengths. We then subtract the relative final transmission spectrum from the individual relative spectra, and display these transit depths for all visits in each channel. The relative transit depths are consistent across the 10 visits with the final relative spectrum. Points that are significantly away from zero (dotted lines) have larger error bars which makes them less important in the weighted average and thus do not bias our transmission spectrum. The uncertainty on the combined spectrum is also displayed as the grey region. In the top left panel, the broadband transit depths are shown, centered on the average across the 10 visits, highlighting the variability in the observed broadband transit depths from the different visits.} \label{fig:comparison}
\end{figure*}

Each of the 11 HST/WFC3 transit observations consisted of $\sim$3 hours of observing time spanning three telescope orbits with $\sim$1~hour gaps between them. Each transit observation is thus composed of one orbit before, during, and after transit. The transit time series were obtained with the G141 grism using the spatial scan mode. In order to optimize the duty cycle of our observation, we used both the forward and backward detector scans. We discard one of the transits from our analysis (November 1st 2019) since a pointing maneuver cut orbit 1 short before the ramp was stabilized, effectively carrying an unusually strong ramp to orbit 2 and polluting the in-transit observations.

% Old longer version
%We discard one of the transits from our analysis (November 1st 2019) since a pointing maneuver of the observatory cut the first orbit short, ending the exposure before the characteristically strong ramp of orbit 1 was stabilized. This carried over to the second orbit in the form of a stronger ramp than expected, effectively polluting the observations where the target was transiting and preventing us from extracting useful scientific information from this visit.

We reduce the observations following standard procedures for HST/WFC3 observations \citep[details in ][]{benneke_sub-neptune_2019, benneke_water_2019}. In order to minimize the background contribution, we subtract consecutive reads up-the-ramp and then add together the background subtracted frames. We then construct flat-fielded images from the flat-field data product provided by STScI. We use a normalized row-added flux template in order to remove and replace outlier pixels in our frames. We follow \citet{benneke_sub-neptune_2019} in order to correct for the slight slanted shape of the trace on the detector, which is introduced by the spatial scan mode, using a trapezoidal shape integration scheme for the wavelength bins, which we choose to be 30\,nm wide. Our flux integration does not perform presmoothing and does account for partial pixels along the trapezoidal bin boundaries. Finally, in order to account for the small drift of the star across the detector during the observations, we account for a small position shift which is measured in each frame.

\section{Data Analysis}\label{sec:data}
We perform the light-curve fitting of our 10 transits of GJ~9827\,d individually using the ExoTEP framework \citep{benneke_spitzer_2017, benneke_sub-neptune_2019, benneke_water_2019}. We use ExoTEP to jointly fit the transit model with a systematics model and a photometric noise parameter in a Markov Chain Monte Carlo scheme \citep{foreman-mackey_emcee_2013}. We decide to fit the transits individually since they display variability in the transit timings (see Figures \ref{fig:wlc}, \ref{fig:TTVs}, and Section \ref{sec:variable}).

Each visit in our data set consists of three HST orbits which do not cover the full transit duration of GJ~9827\,d (Figure \ref{fig:wlc}). The in-transit observations only occur during the second orbit of each visit, either observing the ingress, the middle of transit, or the egress (Figure \ref{fig:wlc}). For that reason, we cannot obtain reliable constraints on the orbital parameters out of the partial transit observations and decide to use a fixed orbital solution during the fits ($b$=0.91, $a/R_\star = 19.88$). 

Since the visits do not include a burn-in orbit, we cannot follow the standard procedure to discard the first orbit, which displays a stronger ramp in time as the detector is still settling \citep[e.g.,][]{kreidberg_clouds_2014}. We rather choose to keep the 2-4 last points of orbit 1 in each visit (Figure \ref{fig:wlc}), as the strong ramp has stabilized by then, and it provides essential baseline information, especially for visits that only have mid-transit or egress data in orbit 2 (Figure \ref{fig:wlc}).  For all orbits, we discard the first forward and backward scans.

Because GJ~9827 is a close-in, near-resonant system, some of the visits in our data set also include transits of GJ~9827\,b. Given the partial coverage of our visits, we simply remove the points where GJ~9827\,b is expected to transit, which affects visits 5 and 10. Visits 6 and 7 also include a transit of planet b, but it happens in the first orbit which is already mostly discarded. Finally, we remove the last 5 points of visit 3 since they are clear outliers. 

\subsection{White-light-curve fit}
% \subsubsection{HST/WFC3 systematics model}
We fit for systematics trends in the normalized transit light curves simultaneously with the transit model using an analytical model that allows for a linear slope throughout the visit duration and an exponential ramp in each orbit. Following previous work \citep[e.g., ][]{kreidberg_clouds_2014, benneke_water_2019} we use the following parameterization to account for these systematics:
\begin{equation}
S_{\mathrm{model}}(t)=(c\,S(t)+v\,t_v ) \times (1 - e^{(-a\,t_{\mathrm{orb}}-b - d)}).
\end{equation}
Here, $c$ is the normalization constant, $v$ is the linear slope throughout the visit, $a$ and $b$ are the rate and amplitude of the exponential ramp in each orbit, and $d$ is an offset only for first orbit reads. $S(t)$ is a function that is equal to 1 for forward scans and to $s$ for backward scans, allowing to correct for the offset between forward and backward scans. Finally $t_v$ is the time since the start of the visit, while $t_{orb}$ is the time since the start of the orbit. 

We produce our transit light-curve models using the Batman package \citep{kreidberg_batman_2015}. Since we are not trying to fit the orbital solution, the only two astrophysical parameters that we are fitting in the transit light curve are the transit depth, and the mid-transit time (Figure \ref{fig:TTVs}). For the limb darkening, we use the Exotic-LD package \citep{david_grant_2022_7437681} to compute the coefficients using 1D stellar models \citep{kurucz1993atlas9} and the quadratic limb darkening law. The impact parameter and semi-major axis are set to the values in \citet{niraula_three_2017}. We obtain the posterior distribution on our parameters by running a MCMC analysis individually for each of the 10 visits. We use four walkers per parameter and all priors are uniform. The 10 white-light-curve fits are shown in Figure \ref{fig:wlc}.

% \begin{figure*}[!tbp]
%   \centering
%   \subfigure{\includegraphics[width=0.48\textwidth]{wfc3_wlc.pdf}}
%   \hfill
%   \subfigure{\includegraphics[width=0.48\textwidth]{wfc3_wavelc.pdf}}
%   \caption{White-light-curve fit (left) and a typical spectral-light-curve fit (right) from the joint analysis of the eight WFC3 transit observations of K2-18b. The top panel shows the best-fitting model light curves (black curve), overlaid with the systematics-corrected data (circles). Residuals from the light-curve fits are shown in the middle panels. The bottom panels shows a histogram of the residuals normalized by the fitted photometric scatter parameter for each respective transit. The residuals follow the expected Gaussian distribution for photon noise limited observations.}
%   \label{fig:HST}
% \end{figure*}

\begin{table}[t!]
\centering
\begin{tabular}{lccc}
\hline
\hline
Instrument               &  Wavelength            &  Depth       & $\pm$1$\sigma$ \\
                         &  [$\mu$m]              &  [ppm]       & [ppm]\\
\hline
HST/WFC3                 &  1.10 -- 1.13    &  941.3     & 59.7 \\
                         &  1.13 -- 1.16    &  1003.3     & 28.3 \\
                         &  1.16 -- 1.19    &  939.6     & 24.5 \\
                         &  1.19 -- 1.22    &  945.5     & 23.4 \\
                         &  1.22 -- 1.25    &  987.7     & 22.4 \\
                         &  1.25 -- 1.28    &  965.7     & 18.8 \\
                         &  1.28 -- 1.31    &  921.0     & 22.3 \\
                         &  1.31 -- 1.34    &  998.7     & 20.3 \\
                         &  1.34 -- 1.37    &  1018.9     & 22.1 \\
                         &  1.37 -- 1.40    &  986.8     & 21.9 \\
                         &  1.40 -- 1.43    &  1015.4     & 21.2 \\
                         &  1.43 -- 1.46    &  960.4     & 22.1 \\
                         &  1.46 -- 1.49    &  982.5     & 23.0 \\
                         &  1.49 -- 1.52    &  992.2     & 22.5 \\
                         &  1.52 -- 1.55    &  953.8     & 21.3 \\
                         &  1.55 -- 1.58    &  935.3     & 22.8 \\
                         &  1.58 -- 1.61    &  919.3     & 23.5 \\
                         &  1.61 -- 1.64    &  944.0     & 19.1 \\
                         &  1.64 -- 1.67    &  941.2     & 21.9 \\
\hline
\hline
\end{tabular}
\caption{\label{tab:trans_sp} HST/WFC3 near-infrared combined spectrum of GJ~9827\,d}
\end{table}

% \subsection{Spectroscopic light-curve fitting} \label{sec:tr_specfit}
\subsection{Spectroscopic light-curve fit}
We use the white-light-curve fits results in terms of the systematics model to pre-correct the light curves in each spectral bin \citep[divide-white method;][]{stevenson_transmission_2014}. Thus, we divide the spectroscopic light curves by the white-light-curve best-fitting systematics model before starting the fitting. We produce our spectroscopic transit light-curve models similarly as in the white-light-curve case, but we now keep the mid-transit time fixed to the best-fitting value found by the white-light-curve fit. The limb darkening is again modelled with Exotic-LD, and this time, our systematics model is a three-parameter linear slope with trace position (measured during the data reduction, see Section \ref{sec:obs_red}). We thus obtain 10 transmission spectra, one from each visit, by running an MCMC analysis on each. We again use four walkers per parameter and uniform priors on all parameters.

\begin{figure*}[t!]
\begin{center}
\includegraphics[width=0.52\linewidth]{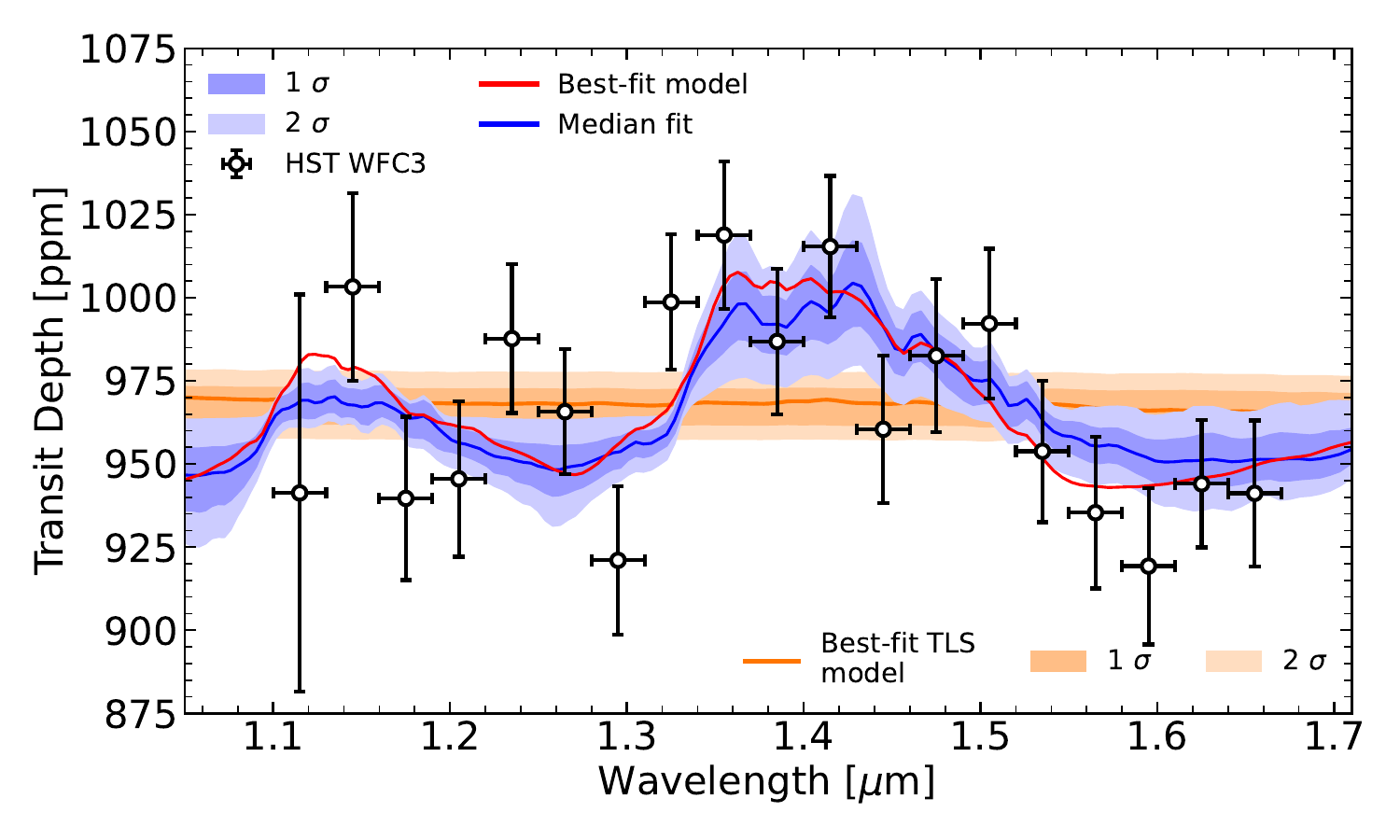}
\includegraphics[width=0.46\linewidth]{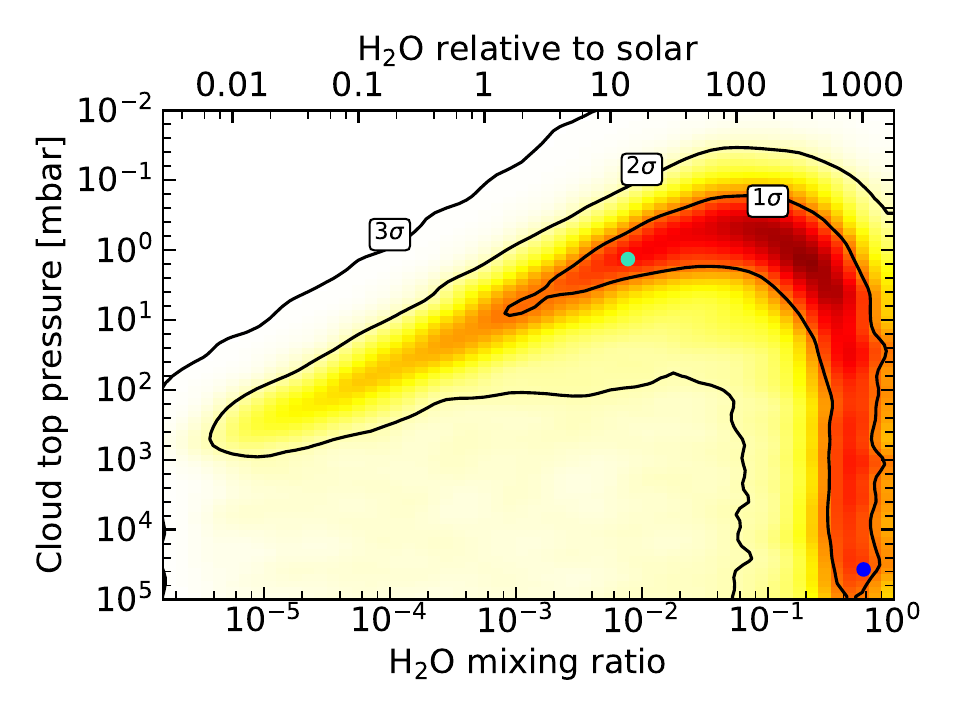}
\end{center}
\caption{Water detection in the transmission spectrum of GJ~9827\,d. \textbf{Left:} Transmission spectrum of GJ~9827\,d (black points) shown with our model transmission spectra constraints from the nested sampling atmosphere retrieval (blue) and from the photometry-informed ``transit light-source effect'' retrieval (orange). The dark blue and light blue shaded regions show the 1$\sigma$ and 2$\sigma$ Bayesian credible intervals  from the atmosphere retrieval respectively. The atmospheric median transmission model is shown in blue and the best-fitting model is shown in red. The best-fitting TLS model is shown in orange along with the 1$\sigma$ and 2$\sigma$ Bayesian credible intervals in light orange. \textbf{Right:} Joint constraints on the cloud-top pressure versus the water mixing ratio derived from our Scarlet well-mixed retrieval. The colored shading describes the normalized probability density as a function of the water mixing ratio (assuming uniform vertical profiles) of the atmosphere, and of the cloud-top pressure. The black contours highlight the 1, 2 and 3$\sigma$ Bayesian credible regions. The water abundance relative to a solar composition atmosphere is shown on the top axis. The posterior probability distribution allows for multiple atmospheric scenarios ranging from H$_2$/He envelopes with small amounts of water to water-dominated envelopes. The blue points identify two representative samples of these two scenarios which are displayed in Figure \ref{fig:spectrum_forward}.}
\label{fig:spectrum_retrieval}
\end{figure*}

\subsection{Combining all the visits together}
We compute a weighted average of our 10 individual transmission spectra to obtain our final transmission spectrum of GJ~9827\,d. In order to verify the robustness of our spectrum, and to ensure that it is not affected by that variability in the observations, we compare the relative transit depth in each channel, for each visit (Figure \ref{fig:comparison}). From each spectrum, we subtract the weighted average (across wavelengths) of said spectrum, essentially making it a relative transit spectrum centered around zero. We then subtract the relative combined spectrum of all visits (also centered around zero) from each individual spectrum to effectively center each spectroscopic transit depth around zero. We then inspect this relative transit depth for each spectroscopic bin and for each visit, in order to ensure that the points in our final transmission spectrum are not affected by outliers (Figure \ref{fig:comparison}). We find that for each spectroscopic channel, all visits mostly agree with the weighted average within error bars, and points that are inconsistent have larger error bars, which makes them much less important in the weighted average, since the weight of each points is inversely proportional to the uncertainty squared (Figure \ref{fig:comparison}). The final average transmission spectrum is presented in Table \ref{tab:trans_sp} and in Figure \ref{fig:spectrum_retrieval}. We decide to discard the last spectroscopic channel (1.67-1.70\,$\mu$m) since it is systematically lower than the rest of the spectrum, and is near the edge of the trace on the detector where the data is less reliable. 

% Keep this for the Discussion section (to shorten)
%Typically, the last points of each orbit are the most constraining as they are much less affected by the exponential ramp, explaining why early-orbit data points lead to looser constraints, and why some visits produce smaller error bars than others. A more thorough discussion of the variability observed in this program is presented in Section \ref{sec:variable}.

%Transit 7 always presents a larger error bar than for the other visits (Figure \ref{fig:comparison}); this is due to the fact that the transit in visit 7 occurred $\sim$1\,hr before the expected transit time, leaving only four early-orbit data points in the egress of the transit.

\section{Atmospheric Modeling}\label{sec:atmosModel}
We perform atmosphere retrievals on our GJ~9827\,d transmission spectrum using the SCARLET framework \citep{benneke_atmospheric_2012,benneke_how_2013,knutson_featureless_2014,kreidberg_clouds_2014,benneke_strict_2015,benneke_sub-neptune_2019, benneke_water_2019, pelletier_where_2021, roy_is_2022}. SCARLET parameterizes the molecular abundances, the cloud deck pressure and the temperature to fit our spectrum. SCARLET uses a Bayesian nested sampling analysis \citep[with single ellipsoïd sampling;][]{skilling_nested_2004, skilling_nested_2006} to obtain the posterior probability distribution of our parameter space and the Bayesian evidence of our models. 

\begin{figure*}[t!]
\begin{center}
\includegraphics[width=0.95\linewidth]{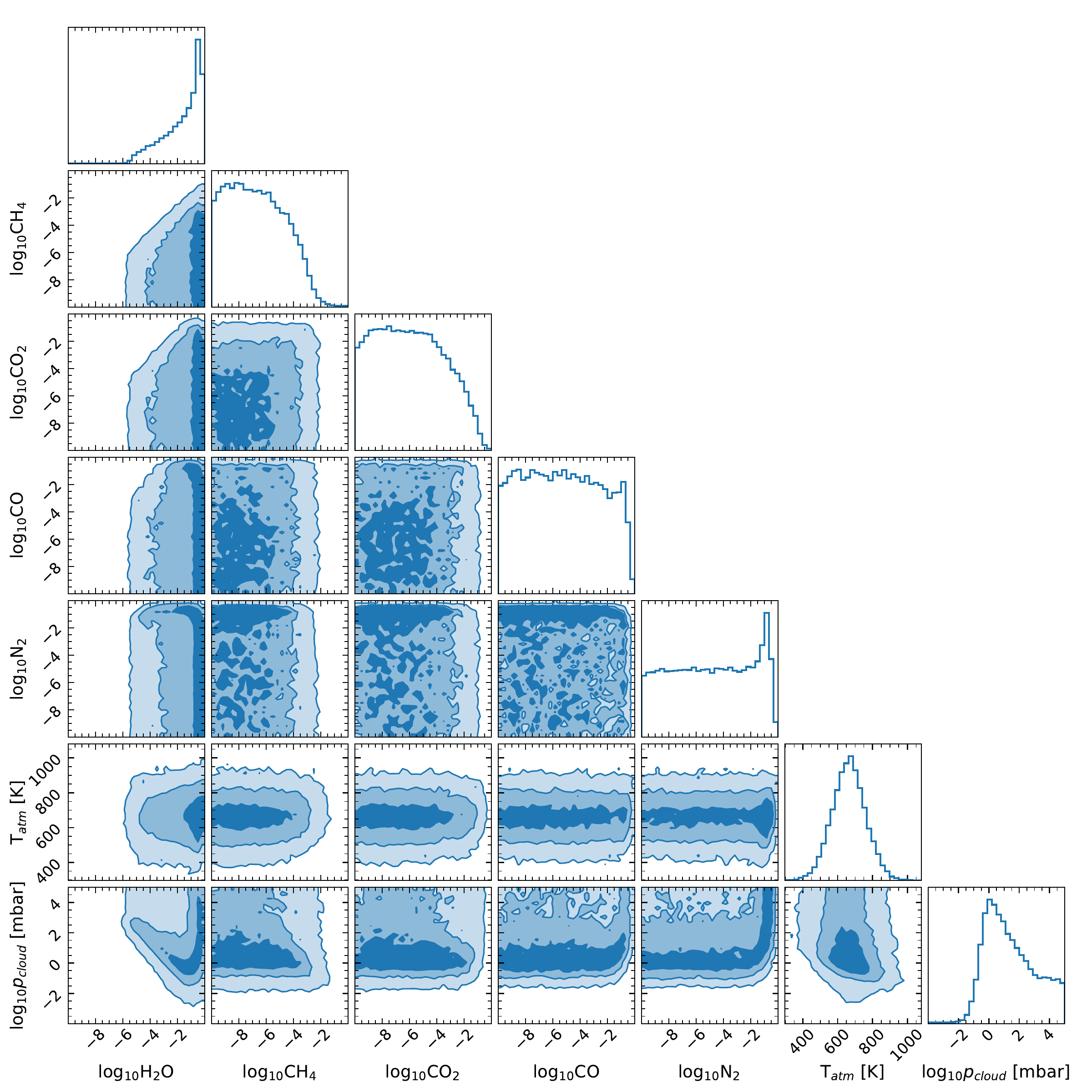}
\end{center}
\caption{Posterior probability distributions of the free parameters used in the SCARLET free chemistry retrieval. The diagonal panels show the marginalized probability distributions of all individual parameters, whereas the off-diagonal panels show the marginalized probability distributions for each pair of parameters as colored shading. The 1, 2, and 3$\sigma$ contours are shown in the 2D posteriors. Water is the only molecule detected in our retrieval analysis.} 
\label{fig:corner}
\end{figure*}

For each set of parameters, SCARLET produces a forward atmosphere model in hydrostatic equilibrium \citep{benneke_strict_2015}, computes the opacity associated to each molecule throughout the 40 vertical (pressure) layers of the model, computes the transmission spectrum for that model and finally performs the likelihood evaluation. The model transmission spectra produced at each step have a resolution of 16000 and are then convolved to the wavelength bins of the observed spectrum. The molecules considered in our retrieval are H$_2$O, CH$_4$, CO$_2$, CO and N$_2$, as well as H$_2$ and He which are not parameterized and rather fill up the atmosphere \citep{benneke_how_2013}. 

\begin{table}[t!]
\centering
\begin{tabular}{lcccc}
\hline
\hline
Retrieval model &  Evidence & Bayes Factor &  N$_\sigma$ \\
                &  ln(Z$_i$)    & B = Z$_{ref}$/Z$_i$& \\ 
\hline
All molecules & -90.68 & ref. & ref.\\
+ clouds & & & \\
H$_2$O removed & -94.96 & 72.52 & 3.39 \\
CH$_4$ removed & -90.24 & 0.65 & 0.90 \\
CO$_2$ removed & -90.45 & 0.79 & 0.90 \\
CO removed & -90.60 & 0.92 & 0.90 \\
N$_2$ removed & -90.73 & 1.06 & 1.14 \\
Clouds removed & -90.78 & 1.10 & 1.23 \\
H$_2$O, CH$_4$ removed & -94.88 & 66.92 & 3.36 \\
Flat spectrum & -97.11 & 620.76 & 4.01 \\
\hline
\hline
\end{tabular}
\caption{\label{tab:detect} Bayesian model comparison results from our SCARLET atmosphere retrievals in the free chemistry setting}
\end{table} 

We assume well-mixed vertical chemical profiles, where the abundances of molecules do not vary throughout the atmosphere. We choose a log-uniform prior on the abundance of each molecule ranging from 10$^{-10}$ to 1 in volume mixing ratio.  We assume a constant temperature structure throughout the atmosphere and use a Gaussian prior centered on the planet's equilibrium temperature \citep[680\,$\pm$\,100\,K;][]{rodriguez_system_2018} on that parameter. The parameterization also includes a cloud deck top pressure, which blocks all light rays going through that pressure level. We again use a log-uniform prior on that parameter from 10$^{-4}$ mbar to 10$^5$ mbar. Thus, our atmosphere retrieval includes seven free parameters in total (or less when molecules are removed, see Table \ref{tab:detect}).
%, as it otherwise is fully unconstrained and collapses to the lower bound of the uniform prior

\begin{figure*}[t!]
\begin{center}
\includegraphics[width=0.65\linewidth]{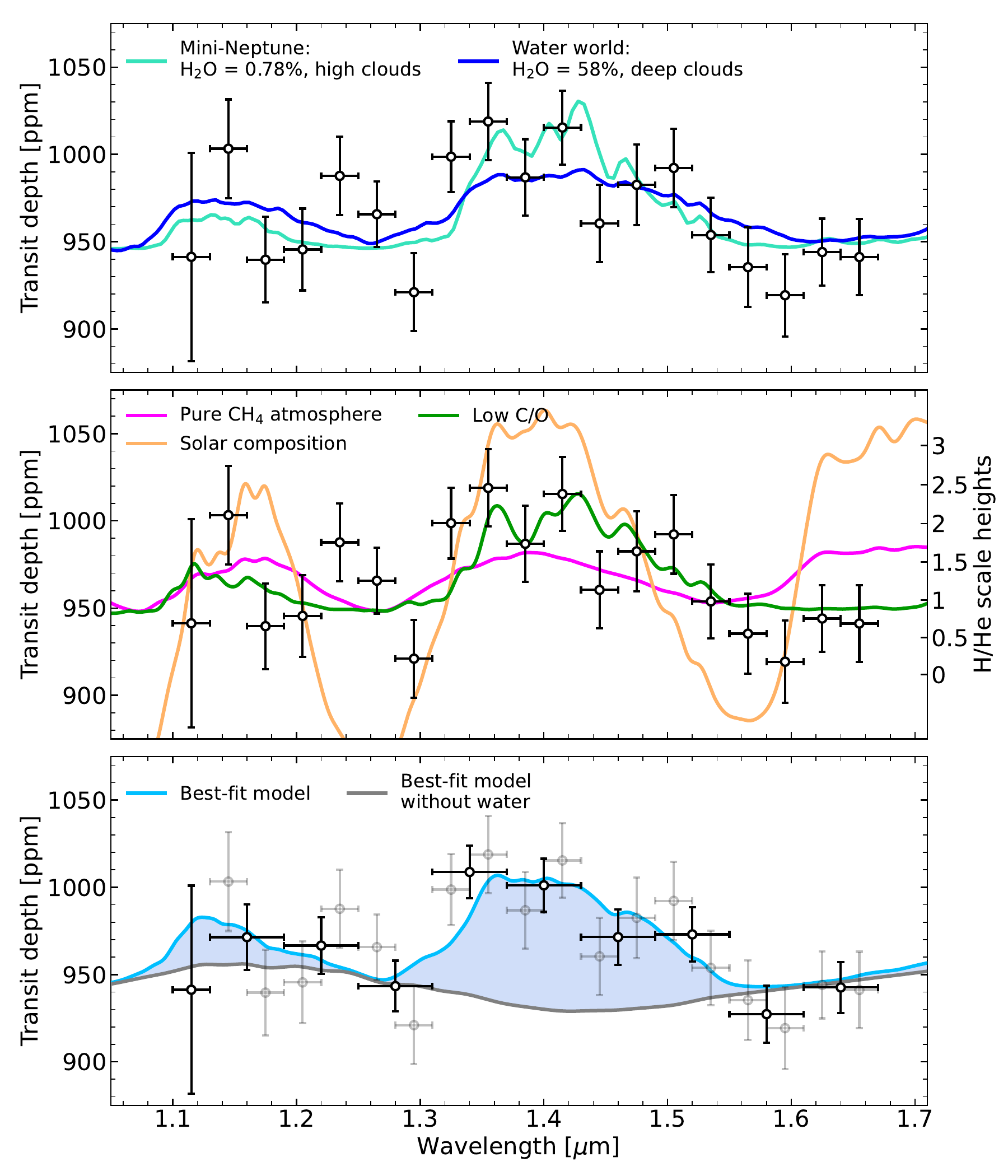}
\end{center}
\caption{HST/WFC3 transmission spectrum of GJ~9827\,d (data points) along with SCARLET forward atmosphere models (colored lines). \textbf{Top:} Two samples from our well-mixed retrieval analysis (Figure \ref{fig:spectrum_retrieval}) are shown, representing the mini-Neptune scenario with a cloudy H$_2$/He atmosphere composed of $\sim$1\% water (pale blue) and a water world scenario with a water-rich atmosphere (dark blue). \textbf{Middle:} A secondary atmosphere model for a pure methane envelope is also shown (red) in order to highlight the methane absorption features. Chemically consistent models (still assuming a uniform temperature profile) are shown for a cloud-free, solar composition case (solar metallicity, solar C/O; orange) and for a cloudy case with C/O=0.1  and a 100\,$\times$\,solar metallicity (green). The observed spectrum is inconsistent with cloud-free low-metallicity scenarios and prefers water absorption features to methane absorption features, mainly around 1.2 and 1.65\,$\mu$m. The strength of the features in the spectrum is also displayed in units of H/He scale heights (right axis). \textbf{Bottom:} The best-fit model from the retrieval analysis is shown (pale blue), along with the transmission spectrum of the same model once the water opacity is turned off (grey). The contribution of water opacity to the spectral signatures is highlighted in blue. We also present a binned version of the transmission spectrum where points are binned together by pair with the exception of the blue-most channel.}
\label{fig:spectrum_forward}
\end{figure*}

\section{Results}\label{sec:resu}
The observed transit spectrum of GJ~9827\,d displays a water absorption feature at 1.4\,$\mu$m (Figure \ref{fig:spectrum_retrieval}). Qualitatively, the transit depths in the spectrum are deeper in the 1.4 $\mu$m water band followed by a downward slope that follows the wing of the water absorption feature. Quantitatively, a Bayesian model comparison analysis of our well-mixed retrievals \citep{benneke_how_2013} prefers models that include the presence and absorption of water with a significance of 3.39$\sigma$ (Bayes Factor = 72.52; Table \ref{tab:detect}) compared to models that do not include water.

\subsection{Metallicity-clouds degeneracy}
Our free chemistry retrievals show that the data can both be explained by a water-rich atmosphere, where water is the most abundant molecule, as well as with a H$_2$/He-dominated atmosphere that still contains a small amount of water (Figure \ref{fig:spectrum_retrieval}). At 1$\sigma$, models with a water mixing ratio between 0.02\% and 80\% are preferred by the spectrum. When compared to the amount of water in a solar metallicity envelope, we see that this interval in abundance ranges from 1 $\times$ solar metallicity models, which are dominated by H$_2$ and He gas, to 1000 $\times$ solar metallicity models, where water is now the principal species (Figure \ref{fig:spectrum_retrieval}). 

In the cases where water is present in small amounts, a cloud deck is needed to explain the observed transit spectrum (Figures \ref{fig:spectrum_retrieval}, \ref{fig:corner}). This is due to the fact that the observed spectrum does not display the large amplitude expected from cloud-free models (Figure \ref{fig:spectrum_forward}). This need for clouds in low mean-molecular-weight atmospheres is also seen in the marginalized probability distributions of the other molecules (Figure \ref{fig:corner}), since their spectral features are inconsistent with the observed spectrum, and thus must be muted by clouds in order to yield high-likelihood models. In cases where the water is more abundant and becomes the principal molecule, the spectra naturally display muted features because of the high density of the atmospheres and lower atmospheric scale height, thus removing the need for high clouds. In this water-rich scenario, the constraint on the cloud-top pressure disappears, explaining the observed posterior distribution (all deep cloud decks become equally consistent; Figure \ref{fig:spectrum_retrieval}).

\subsection{Upper limits on other molecules}
While methane is known to display a similar 1.4\,$\mu$m absorption feature as water \citep{bezard_methane_2022}, it remains disfavored in our free chemistry retrieval analysis. The spectral signatures of methane not only include a feature around 1.4\,$\mu$m, but also at 1.2\,$\mu$m and at 1.7\,$\mu$m, as shown by SCARLET forward atmosphere models for a pure methane envelope and for a solar composition atmosphere (Figure \ref{fig:spectrum_forward}). However, the observed transmission spectrum of GJ~9827\,d does not display these absorption features at 1.2 and 1.7\,$\mu$m (Figure \ref{fig:spectrum_forward}) and is in better agreement with the water models that display a smaller feature at 1.2\,$\mu$m, no absorption at 1.7\,$\mu$m, and a broader feature at 1.4\,$\mu$m. In order to obtain chemically consistent models that agree with the observed transit spectrum, the carbon-to-oxygen (C/O) ratio must be decreased in order to favor O-bearing molecules (here, water) and a cloud deck must be included to mute the spectral amplitude of the water absorption features. Such models (e.g. C/O = 0.1, Metallicity = 100 $\times$ solar and pCloud = 1\,mbar) yield qualitatively and quantitatively similar transmission spectra to those favored by our free chemistry retrieval (Figures \ref{fig:spectrum_retrieval}, \ref{fig:spectrum_forward}).

No other molecule besides water is statistically detected by our retrievals (Table \ref{tab:detect}, Figure \ref{fig:corner}). However, we can derive upper limits in their abundances based on our results from the free chemistry retrievals, either from the non-detection of specific spectral features (e.g., CH$_4$) or because too much of anyone species increases the mean-molecular-weight of the atmosphere, eventually yielding a flat spectrum (e.g., N$_2$). Thus, our retrievals allow us to constrain the upper limits on the mixing ratios of CH$_4$, CO$_2$, CO and N$_2$ to 3.04, 19.4, 52.5 and 60.4 \% at 3$\sigma$ significance. 

% Re-phrase to shorten
% As explained above, molecules such as CH$_4$, which do have spectral signatures in HST/WFC3's range, can be ruled out because of the absence of those features in the observed transit spectrum of GJ~9827\,d. Moreover, for spectrally inactive molecules across the HST/WFC3 range, such as N$_2$, we can put upper limits on their abundances since large amounts of anyone species lead to high mean-molecular-weight atmospheres and muted (eventually flat) spectra which are disfavored by the observed spectrum.

%Finally, a flat transmission spectrum showing no molecular absorption is ruled out by the observed spectrum at 4.01 $\sigma$.

\subsection{Significance of a featureless spectrum}
In order to evaluate how our spectrum deviates from a featureless spectrum, we compute the deviation from the best-fitting straight line using $\chi^2$ statistics. Using the binned version (bottom of Figure \ref{fig:spectrum_forward}), we obtain that the transmission spectrum of GJ~9827\,d deviates from a straight line at 3.24$\sigma$. In order to assess how our water detection compares to a featureless flat spectrum within the Bayesian paradigm, we compute the Bayesian evidence of a one-parameter flat line model $\mathcal{Z}_\textrm{flat}$. Given the simplicity of this one-parameter model, we do not need to use SCARLET nested sampling to obtain that value, and rather numerically estimate it via the following analytical solution: 
\begin{equation}
    \begin{aligned}
        \mathcal{Z}_\textrm{flat} &= \left(\frac{1}{\sqrt{2 \pi}}\right)^N \times \left(\prod_{i=1} ^N \frac{1}{\sigma_i}\right) \times \left(\frac{1}{\theta_2 - \theta_1}\right) \\
        & \qquad \qquad \times \int_{\theta_1} ^{\theta_2}  \exp\left(-\sum_{i=1} ^N \frac{(D_i - \theta)^2}{2\sigma_i^2}\right)    d\theta,
    \end{aligned}
\end{equation}
where $N$ is the number of points in the spectrum, $\sigma_i$ is the uncertainty of the $i^{\mathrm{th}}$ point of the spectrum denoted $D_i$, and where $\theta_1$ and $\theta_2$ are the limits of our uniform prior on the transit depth parameter $\theta$. This allows us to show that our atmosphere model is preferred to the flat spectrum model at 4.01$\sigma$ (Bayes Factor = 620.76; Table \ref{tab:detect}).

\subsection{Ruling out stellar contamination}
The Transit Light Source Effect \citep[TLS,][]{rackham_transit_2018} can mimic water features at the 20 ppm level or more for modestly spotted stars under certain observational configurations, and could thus create the feature observed in our transmission spectrum. The best constraint available for the starspot coverage and contrast for GJ 9827 comes from the K2 Campaign 12 (C12) lightcurve, with an SFF-derived \citep{vanderburg_technique_2014} peak-to-valley amplitude of 0.45\%, slightly lower than the typical 0.7\% for K6 spectral types \citep{rackham_transit_2019}.  Coarse scaling laws can relate the observed K2 amplitude to the surface coverage of starspots, under assumptions of size, number, and location of spots on a stellar surface \citep{rackham_transit_2018,guo_effect_2018}.  For a K6 spectral type, 0.45\% amplitude variations relate (conservatively) to spot-covering fractions of 1-4\% \citep{rackham_transit_2019}. Thus, we adopt a spot contrast typical of a K6 star and a filling factor of 1-4\% \citep{rackham_transit_2019}.  Under these assumptions, a planet with a 1\% transit depth could expect an H$_2$O contamination of $<$15 ppm from unocculted spots.  GJ 9827\,d's much smaller transit depth of $<0.1\%$ would therefore yield a negligible TLS water contamination of $<1.5$ ppm. 

We perform a retrieval on our combined transit spectrum in which we fit for TLS parameters rather than for atmospheric parameters. We use the 1-4\% spots filling factor as a prior in this TLS retrieval, and the other parameters are the photospheric temperature, the spots' temperature difference and a scaling factor. Our TLS models simulate unocculted star spots by averaging the stellar spectrum of the star's photoshere with the spectrum of the cooler spots (weighted by the spots coverage) based on Phoenix stellar models \citep{husser_new_2013}. It then simulates the transit of an airless planet to obtain the stellar-contaminated transit spectrum.  When using the photometry-derived prior, we find that the TLS fitting is restricted to flat models which indicate that stellar contamination is not expected for that system (Figure \ref{fig:spectrum_retrieval}). 

Repeating the same TLS retrieval with nonrestrictive uniform priors on all parameters further demonstrates that the signal cannot be caused by the star. We run the same TLS retrieval as described above, but without using the 1-4\% spots filling factor prior, to see under what stellar conditions the signal can be explained by the star. We find that, in order for the TLS to reproduce the signal in our spectrum, not only does the spots parameters need to be unrealistically large (73\% spots coverage and $< -794\,$K spot temperature difference), but the model needs to adopt a strong positive ramp towards short wavelengths \citep[as in][]{moran_high_2023}, which becomes inconsistent with the K2 transit depth measurement. We thus conclude that stellar contamination cannot explain the feature in the transmission spectrum of GJ 9827\,d, and that the water detection comes from the planet's atmosphere.

% %Old discarded paragraph 
%The detection of a water absorption feature is also highlighted in our chemically consistent retrieval, which allows for atmospheres with metallicities ranging from 1 to 1000\,$\times$\,solar, provided a cloud deck is present to attenuate the feature strength. Under the chemical equilibrium assumption, the retrieval still prefers atmosphere models where water is the only detectable absorbing species. However, given the more rigid parameterization, this is manifested in a C/O posterior that collapses to the lower bound allowed by our uniform prior, so that oxygen-bearing species like H$_2$O dominate over carbon-bearing species. It can also be seen in the presence of clouds that are going to mute the smaller features of other molecules such as CH$_4$ and only leave the larger water feature to be detectable above the cloud deck. It also affects the shape of the posterior at high metallicities since in that regime, multiple high-mean-molecular-weight species become abundant, and this decreases the atmospheric scale height and destroys the absorption features more effectively than in the well-mixed case, where the water abundance can be increased independently of the other molecules.

% %Old figure idea
% \begin{figure}[tb]
% \begin{center}
% \includegraphics[width=\linewidth]{placeholder_populationFigure.png}
% \end{center}
% \caption{TO BE UPDATED. Some sub-Neptune population figure similar to this. Could actualize this figure with GJ~9827d and new planets. } \label{fig:Population}
% \end{figure}

\section{Discussion and Conclusions}\label{sec:disc}
The HST/WFC3 transmission spectrum of GJ~9827\,d presented in this work provides a precious target in the population of sub-Neptune exoplanets for which we have precise transmission spectra; and highlights GJ~9827\,d as the smallest exoplanet with an unambiguous atmospheric molecular detection to date. Compared to the other characterized sub-Neptunes, our detection of a $\sim$1 H/He scale height water feature (Figure \ref{fig:spectrum_forward}) makes it stronger than for similarly hot sub-Neptunes, although it remains broadly consistent with the previously observed trend where hotter sub-Neptunes display stronger H$_2$O amplitudes \citep{crossfield_trends_2017}. Our analysis of the 10 observations of GJ~9827\,d's transits also revealed some variability in this planet's orbit, which is to be considered for further monitoring of the system with state-of-the-art facilities such as JWST. Finally, our detection of a water feature in GJ~9827\,d's transit spectrum provides the first detection of water in the atmosphere of a potential water world, which, when combined with GJ~9827\,d's large mass-loss rate, provides a first line of evidence for this sub-Neptune hosting a water-steam dominated atmosphere.

\subsection{Variability in the transits of GJ~9827\,d}\label{sec:variable}
The analysis of the 10 transits of GJ~9827\,d with HST revealed a significant variability in the transit timings observed from one visit to the other (Figure \ref{fig:TTVs}). While this variation is not surprising for a near-resonant system and did not impact the features observed in the transit spectrum (as shown by our analysis of the relative spectra; Figure \ref{fig:comparison}), it still is in contrast with the previously observed TTVs for this system which were of the order of $\sim$3 minutes \citep{niraula_three_2017}. However, the 5-10 minutes TTVs observed for planet d in this work are consistent with an independent study of the TTVs of the GJ~9827 system combining all photometric and radial velocity data (Livingston et al., in prep.).

%Some visits where the second HST orbit collected data only around the mid-transit time yield large uncertainties in the transit timing, as there is no ingress or egress to anchor the start and end of the transit. While a thorough photodynamical TTV analysis of this system is out of the scope of this work, it should be characterized in more details in the near-future. 

%and could potentially reveal a yet-unknown outer planet that could cause these variations. 
%However, other visits, such as visit 7, consistently retrieve transit timings that are $\sim$1\,hr away from the expected transit time, no matter the transit model and the orbital parameters chosen for the fit.

The limited number of in-transit data points in the time series in this program could also explain the range of transit depths observed in our results. As described earlier, each HST orbit displays an exponential ramp in time that is fitted by our systematics model. This ramp has a much stronger effect in the first few integrations than in the last few integrations of each orbit, when it has settled. Thus, HST/WFC3 observations inherently provide better quality observations towards the end of each orbit. When considering the individual visits in our program, it seems that egress visits yield deeper transit depths than mid-transit or ingress visits (Figures \ref{fig:wlc}, \ref{fig:comparison}). This could then be explained by the fact that the different visits in our data set have varying in-transit data quality depending on whether the late-orbit integrations are in the transit (ingress and mid-transit visits) or are in the baseline (egress visits). For instance, visit 6, which is an egress observation, displays a deeper transit depth than in the other visits. However, the relative shape of the transmission spectrum, which is the quantity of interest for this work, is consistent with the other visits (Figure \ref{fig:comparison}). 

%While this explanation provides a possible reason for some of the observed variability in our program, it remains in the realm of small number statistics and continuous monitoring of multiple full transits of GJ~9827\,d will be needed in order to obtain an updated orbital solution for the planet. 
%Visit 7 demonstrates this effect clearly, as it only displays 4 early-orbit in-transit data points, which leads to the large uncertainty on the retrieved parameters from that visit. 

Another potential source of the TTV and transit depth variability observed in our program is star-spot crossing. GJ~9827 has been shown to display quasi-periodic flux ($\sim$0.45\%) variations with a period of $\sim$30 days \citep{rodriguez_system_2018,teske_magellanpfs_2018, prieto-arranz_mass_2018, rice_masses_2019, kosiarek_physical_2021}. If these stellar variations were caused by stellar spots, then using a fixed orbital solution and a transit model that does not include the effect of spots in our light-curve fits could lead to biases in our retrieved parameters. 

% However, the partial transit coverage of each individual transit prevents us from including stellar spots in our fits, and we have shown that we expect a spots filling factor of 1-4\% for this system. This issue is likely to remain unsolved until, again, repeated transit time-series of GJ~9827\,d are obtained with a high-precision high-cadence continuously-observing facility such as JWST, which would provide a detailed uninterrupted light curve covering the full transit. 

% If these stellar variations were caused by stellar spots, it could provide an alternative explanation to the variable transit signals observed in our work. With a large impact parameter, GJ~9827\,d transits near the outer annulus of the star \citep{rodriguez_system_2018}, where its projected area on the star is made larger by the 3D stellar surface, potentially making it more likely to hide stellar spots or regions of stellar activity.

Because of the variability discussed above, we decided to fix the orbital solution and limb darkening coefficients for the light-curve fits in our program (Section \ref{sec:data}). In order to ensure that the limb darkening coefficients and stellar parameters chosen do not affect our atmospheric inference, we repeat our light-curve fits for multiple assumptions on the limb darkening. Using quadratic limb darkening coefficients, we reproduce the same fitting but using a 3D stellar model \citep{magic2015stagger} when computing the coefficients. We further try the light-curve fits by varying the effective stellar temperature to the +1$\sigma$ and $-1\sigma$ values of that parameter \citep{kosiarek_physical_2021}. In all cases, we find that the limb darkening and choice of stellar parameters only affect the retrieved spectrum with a constant offset throughout the wavelength range, and that the relative spectra all show the water absorption feature and are all consistent within 1$\sigma$. This thus shows that our choice for the stellar and limb darkening parameters does not affect the shape of the transmission spectrum, and subsequently, our atmospheric analysis.

Similarly, given the difficult observational setting, we test the robustness of the spectrum to the systematics models used by trying two alternative light-curve fitting methods. First, we start from the divide-white corrected spectroscopic light curves and jointly fit a relative transmission spectrum. To do so, we jointly fit (across visits) the relative transit depth in each channel, where the broadband average of the individually fitted spectra is subtracted for each visit (since there sometimes are discrepancies between the white-light-curve transit depths, and average spectral depths in HST/WFC3 data). Secondly, we repeat the method presented in Section \ref{sec:data}, but using the RECTE systematics model and orbit 1 \citep[and not using the divide-white method;][]{zhou_physical_2017} for the seven visits which are not affected by transits of planet b in orbits 1 or 2. This gives us 7 transmission spectra that we combine with a weighted average. Both of these methods produce relative transmission spectra that are consistent within uncertainties with the one presented in Table \ref{tab:trans_sp}. The spectrum obtained from the RECTE models has increased uncertainties, both from the smaller number of visits and from increased fitted scatter in some visits, but is still in agreement with the other two spectra. The spectrum presented in this work is thus robust to the choice of systematics model.

\subsection{Water in the envelope of a potential water world}
The water detection in the transit spectrum of GJ~9827\,d makes it the first water world candidate with an atmospheric water detection consistent with a water-rich envelope. It thus positions itself in the sample of potential water worlds, with other small sub-Neptunes and super-Earths such as Kepler-138\,d \citep{piaulet_evidence_2022}, L~98-59\,d \citep{kostov_l_2019}, TOI-1685\,b \citep{bluhm_ultra-short-period_2021}, GJ~3090\,d \citep{almenara_gj_2022}, TOI-270\,d \citep{gunther_super-earth_2019, mikal-evans_hubble_2023}. In contrast, a water feature was also detected in the transit spectrum of TOI-270\,d \citep{mikal-evans_hubble_2023}, but the analysis revealed that the H-rich atmosphere scenario was favored for this sub-Neptune, showing that the line can be fine between a mini-Neptune and a water world.

With its small mass of 3.42\,M$_\oplus$ \citep{kosiarek_physical_2021} and its proximity to its host star (6.2\,d orbit), the estimated mass-loss rate of GJ~9827\,d is $>$0.5\,M$_\oplus$/Gyr \citep{krishnamurthy_absence_2023}. With an estimated age around 6 Gyr \citep{kosiarek_physical_2021}, GJ~9827\,d is unlikely to retain an extended H$_2$/He envelope today. Furthermore, monitoring of GJ~9827\,d's spectrum in the search of H$\alpha$ and HeI signatures with Keck/NIRSPEC \citep{kasper_nondetection_2020} CARMENES \citep{carleo_multiwavelength_2021} and IRD \citep{krishnamurthy_absence_2023} resulted in no evidence of an extended H$_2$/He atmosphere around the planet. Hence, the H-rich scenario with a smaller water abundance would provide a somewhat contradictory statement to the previous studies that observed GJ~9827\,d from the ground. However, the water-rich scenario can both explain the observed HST transit spectrum, as well as the non-detection of H$\alpha$/HeI lines from ground-based studies. The water-dominated envelope is thus the compositional scenario that explains all of the data at hand on this system in the most natural way. 

% While the other water world candidates are consistent in bulk density with volatile-rich envelopes, the main volatile species remains unknown. We know water to be an abundant and stable molecule in planetary atmospheres, but these potential water worlds could also be enriched by other volatiles such as CH$_4$, CO$_2$, NH$_3$, etc, as the only information on their composition we have thus far comes from their average density. However, the detection of water in GJ~9827\,d's spectrum and the non-detection of other absorption features narrows our search for this sub-Neptune, as the question is to now learn whether water is present in solar-composition amounts in an otherwise H-rich envelope, or if it actually represents the dominant molecule in the envelope. 

In this water-rich scenario, GJ~9827\,d would thus represent a larger, hotter, close-in version of the icy moons of the giant planets in the solar system. Indeed, water is believed to be the dominant volatile of the icy moons of the solar system \citep{schubert_interior_2004}. GJ~9827\,d could then have formed outside of the water ice line, where water ice is available in large amounts as a planetary building block \citep{mousis_jupiters_2019,venturini_nature_2020}. It could then have migrated towards its current stable near-resonant orbit, during which the increasingly important stellar irradiation would have driven an important H$_2$/He loss, and it would be observed today with a high mean-molecular-weight water vapor atmosphere due to its warm temperature \citep{adams_ocean_2008, pierrehumbert_runaway_2023} and its H$_2$/He depletion.

While our transmission spectrum cannot unambiguously distinguish between the H-rich and H-depleted scenarios, we have provided the first water detection in the envelope of a water world candidate, making it a key target for further monitoring with JWST. Transmission spectroscopy of GJ~9827\,d with NIRISS/SOSS and NIRSpec/G395H would provide the high-precision continuous viewing of the full transit of the planet that is needed to explain the variability observed with HST, as well as provide a more precise transmission spectrum that could not only probe the water absorption bands, but also probe for the presence of carbon bearing species like CO and CO$_2$ above 4\,$\mu$m. A JWST transmission spectrum of GJ~9827\,d would thus lift the degeneracy observed in our study (Figure \ref{fig:spectrum_retrieval}) and potentially confirm the water world nature of this sub-Neptune, simultaneously yielding the first direct detection of a water vapor dominated envelope.

\acknowledgements
All of the data presented in this paper were obtained from the Mikulski Archive for Space Telescopes (MAST) at the Space Telescope Science Institute. The specific observations analyzed can be accessed via \dataset[10.17909/dvqh-2r48]{https://doi.org/10.17909/dvqh-2r48}. 

We wish to thank the reviewer for the insightful comments which enhanced the quality of our manuscript. P.-A. R. further thanks L. Bazinet, L.-P. Coulombe and S. Pelletier for their help, comments and ideas throughout the multiple iterations of this work. 

This work is based on observations with the NASA/ESA HST, obtained at the Space Telescope Science Institute (STScI) operated by AURA, Inc. P.-A.R. and B.B. acknowledge financial support from the Natural Sciences and Engineering Research Council (NSERC) of Canada. P.-A.R. further acknowledges support from the University of Montreal, and from the Trottier institute for exoplanets (iREx).  B.B. also acknowledges financial support from the Fond de Recherche Québécois-Nature et Technologie (FRQNT; Québec).

\bibliography{references, others}

\end{document}